\documentclass{emulateapj}

\def\gsim{\;\rlap{\lower 2.5pt
\hbox{$\sim$}}\raise 1.5pt\hbox{$>$}\;}
\def\lsim{\;\rlap{\lower 2.5pt
\hbox{$\sim$}}\raise 1.5pt\hbox{$<$}\;}

\shorttitle{Primordial Gas in LBGs}
\shortauthors{Pan \& Scalo}
\slugcomment{ApJ Letters, in press}

\begin{document}

\title{ Mixing of Primordial Gas in Lyman Break Galaxies}

\author{Liubin Pan and John Scalo}
\affil{Astronomy Department, University of Texas at Austin, Austin, TX\\
panlb@astro.as.utexas.edu; scalo@astro.as.utexas.edu}

\begin{abstract}

Motivated by an interpretation of $z \sim 3$ objects by Jimenez and 
Haiman
(2006), we examine processes that control the fraction of primordial ($Z = 
0$) gas, and so primordial stars, in high-SFR Lyman break galaxies. 
A primordial fraction different from
1 or 0 requires microscopic diffusion catalyzed by a
velocity field with timescale comparable to the duration of star 
formation.
The only process we found that satisfies this requirement for LBGs without 
fine-tuning is turbulence-enhanced mixing induced by exponential stretching 
and compressing of metal-rich ejecta. The time-dependence of the primordial
fraction for this model is calculated. We show that conclusions for 
all the models discussed here are virtually independent of the IMF, including
extremely top-heavy IMFs.
\end{abstract}


\section{Introduction}
Galaxies begin their lives with entirely primordial ($Z=0$) gas. As they
age, metal production and mixing can only reduce the primordial gas 
fraction. We have explored the expected time dependence of the
primordial fraction for various types of mixing and chemical evolution
processes using a kinetic equation for the evolution of the abundance
distribution.
The present Letter addresses the narrower question of whether any 
models for mixing and chemical evolution predict this
transformation occurs at an accessible redshift.

Jimenez and Haiman (2006) \nocite{jim06} (hereafter JH)
showed that several UV properties of a variety of objects, mostly 
Lyman break
galaxies (LBGs), at redshift $z \sim 3$, can all be understood if 
these objects
contain a substantial fraction, about $10$ to $50$ percent, of 
massive stars
with essentially zero metallicity ($Z \lsim 10^{-5}$ $Z_{\odot}$; we 
use $Z$
and metallicity indiscriminantly here). These UV
properties cannot together be explained by a top-heavy IMF, and 
require $Z = 0$ stars; a top-heavy IMF is not required. Massive stars have 
short lifetimes, so their metal abundances
reflect that of the concomitant gas. Thus, if JH are correct, a 
substantial
fraction of the interstellar medium of these galaxies, with star 
formation
ages a few hundred million years, has not been
polluted by any products of nucleosynthesis.\footnotemark\footnotetext{
The existence of $Z=0$ gas requires an IMF deficient in 
stars with $M \lsim 1$ $M_{\odot}$. Otherwise the number of low- 
mass stars
observable today would be large, contradicting observational limits
on the star fraction with very small $Z$ (see Oey 2003) by several 
orders of
magnitude. We discuss the effects of different IMFs below, where we show
that all our conclusions are independent of the IMF as long as it satisfies
the requirement $M \gsim 1$ $M_{\odot}$.}

Motivated by JH and other suggestions for $Z=0$
stars \citep{mal02,shi06}, and the fact that during some early
period in the lives of {\it all} galaxies a transformation from 
primordial to non-primordial must occur, leaving 
spectrophotometric signatures (JH; see Schaerer 2003), we 
examined the viability of a number of models for this transformation. 


A change in the primordial fraction requires microscopic diffusion 
enhanced by
a complex velocity field, such as instabilities in
swept-up supernova shells or a turbulent interstellar medium (ISM),
along with dispersal of nucleosythesis products over many kpc, 
involving a
scale range of $\sim 10^5$. Existing hydrodynamic simulations in
a cosmological context \citep[e.g.,][]{gov04,sca05} therefore cannot
represent the transition of the primordial gas or mixing in
general, since they adopt mixing rules that arbitrarily spread
metals among nearest
neighbor cells or SPH particles. The only hydrodynamic simulations of 
true
mixing of tracers in galaxies were concerned either with turbulent
dispersal of mean metallicity, not mixing \citep{kle03}, or
mixing of initial spatially periodic inhomogeneities by
numerical diffusivity in a turbulent galaxy with no continuing source of
metals \citep{dea02}.  


Instead of simulations, the arguments given here are phenomenological,
in order to clarify the essentials for each physical process. A 
detailed discussion is given elsewhere (Pan and Scalo 2006, hereafter PS)
using a formal probability distribution evolution equation.
Here we report that most processes we examined are
either far too fast or slow to partially erase the primordial fraction 
(Sec. 2), 
except for mixing by turbulence-enhanced diffusivity based on exponential 
stretching of blobs of nucleosynthesis products, 
discussed in Sec. 3.\footnotemark\footnotetext{A recent estimate of the 
effect of SN mixing (sec. 2.4 below) on the primordial fraction at high 
redshift (Tumlinson 2006) apparently used a shell mass much larger than 
found in analytical and numerical calculations of supernova remnant 
evolution 
\citep[see][]{tho98, han06}.}

\section{Estimates of timescales for mixing processes}
\subsection{Star Formation Age}

The existence of a primordial gas fraction $P(t)$ that is not nearly
unity or zero, i.e., both $P(t)$ and $1- P(t)$ are significantly
larger than zero, a condition we refer to as a ``significant'' or 
``intermediate'' primordial fraction, requires a mixing or 
depletion
process whose characteristic timescale is comparable to the star
formation (SF) age, the time since SF began. If the mixing 
timescale
is much smaller, $P(t)$ will
be nearly zero; if it is much larger, $P(t)$ will
remain near unity.
SF ages for LBGs and likely-related objects at somewhat different 
redshifts
have been estimated by Papovich et al. (2001), Shapley et al. (2001),
Erb et al. (2006)\nocite{pap01, sha01, erb06}
and others, using galaxy
evolution models that assume an exponentially decreasing SFR that began
at some time in the past. Although there is much variation, 
median SF ages are $3 \times 10^8$ yr.

These SF ages cannot be significantly different for a number of
reasons. The lookback times of about 11.5 Gyr for z =3 imply a 
strict
upper limit to the SF age of 2 Gyr, and a more likely upper limit of 
1.5 Gyr
(corresponding to $z \sim 8$). The irregular morphologies of the
LBGs \citep{gia02} suggest that LBGs are in the process of formation,
accumulating large fragments of gas and stars through mergers \citep 
{con06}, and probably undergoing one of their first major bursts of SF;
starburst populations have durations, estimated from 
statistics \citep{ken87,nik04}, modeling of integrated light 
features \citep{mar06}, and theoretical 
arguments \citep[see][]{lei01}, that are similar to the estimates in LBGs.
Finally, ages much greater than $3 \times 10^8$ yr would produce  
greater metallicity than observed \citep[see][]{gia02}.

We assume star formation has been a continuous function of time. 
If instead the SFR preceding or during the present episode consists of 
bursts of shorter duration, most of our 
arguments remain unchanged if the SF age is replaced by the accumulated 
duration of SF.

\subsection{Depletion of the Primordial Fraction by Sources}
The JH result of 10-50\% primordial at $z \sim 3$ seems
surprising, but actually galaxies should remain almost completely 
primordial
for billions of years in the absence of microscopic
diffusivity.
SN metal production slowly depletes primordial gas by transferring it 
from a
$Z = 0$ delta function in the $Z$ probability distribution (pdf) to 
another
delta function at a much larger $Z$, the source
metallicity $Z_s$ averaged over the IMF ($\sim$ 0.1 assuming the hot 
ejecta
are well mixed).
Intermediate values of $Z$ cannot be reached without diffusivity.

This suggests the simplest explanation for an intermediate primordial
fraction in LBGs: 50 to 90\% of the primordial gas passed through 
stars that
became SNe. The timescale for this process is the source timescale,
$\tau_{src} = M_{gas}/B R_{SN}$, where $B$ is the star formation rate
and $R_{SN}$ is the returned fraction from SNe averaged over the 
IMF.
We take the total gas mass $M_{gas}$ as
$5 \times 10^{10}$ $M_{\odot}$, extrapolated from gas masses estimated in
the $z \sim 2$ UV-selected sample of Erb et al. (2006).
From a number of studies, we adopt a median SFR of 100 $M_\odot$/yr 
for LBGs at $z \sim 3$ assuming the IMF lower limit 
$M_l=0.1$ \citep{pap01,sha01, gia02, erb06, yan06}.
For this $M_l$, we calculate $R_{SN} \approx 0.1$ for various IMFs using the
ejected masses given in Woosley \& Weaver (1995), Meynet \& Maeder (2002) and
Nomoto et al. (2006)\nocite{woo95,mey02,nom06}, finding
little dependence on metallicity, including $Z=0$, and 20-30\% variation
between studies. Variations in the form of the IMF change the SFR by 
$\sim$50\%, with only a slight effect on $R_{SN}$.
The source timescale is then $\tau_{src} = 5$ Gyr, within a factor of a few
considering the uncertainty in the SFR and $M_{gas}$, too large to 
affect the primordial fraction.

We examined the IMF-dependence of $\tau_{src}$. There are two 
nonstandard IMFs that are especially relevant. 1. Intermediate 
primordial fractions 
require that the IMF lower limit
$M_l \gsim 1$ $M_{\odot}$ to avoid too many $Z=0$ stars observable today.
Such a cutoff does not affect the source timescale: The empirical
SFRs are based on integrated light from massive stars corrected for the rest
of the IMF, and the resulting decrease of the SFR due to the cutoff
is exactly compensated by the increase of the mass ejected by supernovae
per unit mass of stars formed $R_{SN}$. This IMF-independence
of $\tau_{src}$ holds for any cutoff smaller than the lower mass limit
for SNe, $\sim 8$ $M_{\odot}$. By the same argument, such a
cutoff does not overproduce metallicity.  2. A perennially popular 
IMF for $Z=0$ star formation
consists of only very massive stars (VMS) due to the Jeans mass resulting from 
$H_2$ cooling (Hutchins 1976; see Bromm and Larson 2004) although it has been
questioned on a number of grounds \citep{sil06}.
Comparing $H\alpha$ emission per unit SFR for a VMS IMF
(50-500) $M_{\odot}$ of $Z=0$ stars in Schaerer (2003) with the
same quantity for a (0.1-100) $M_{\odot}$ IMF in Kennicutt (1998),
both for a Salpeter IMF (for illustration only),
the SFR for a VMS IMF is 26 
times smaller. Assuming only stars in the range 130-260 $M_{\odot}$ explode 
as pair instability SNe \citep{woo02},
we find $R_{SN}= 0.27$, so $\tau_{src}=50$ Gyr, an order of magnitude 
larger than the normal IMF case.

These results strengthen our conclusion that the formation of massive
stars is far too slow to deplete primordial gas 
significantly in the available time.

\subsection{Filling the Gap by Diffusion}
Without a process to spread metals into the 
``gap'' between
the $Z_s$ peak and the primordial $Z=0$ peak, $P(t)$ would
remain near unity for billions of years.
The only physical process that can fill this gap is microscopic 
diffusivity.
However an estimate of the rate at which diffusivity from 
random sources could
pollute primordial gas in LBGs, using diffusion lengths for the cold 
neutral,
warm neutral, and warm ionized ISM similar to those in 
Oey (2003)\nocite{oey03}, shows that the fraction of the mass of a 
galaxy mixed in time t 
is only $\sim 10^{-5} (t/0.5$ Gyr$)^{5/2}$,
so diffusivity by itself cannot pollute more than a tiny fraction of
the primordial gas over the estimated SF ages.

To reduce the primordial fraction, a velocity field is 
required to catalyze diffusivity. However a velocity field cannot by 
itself affect the global metallicity distribution or primordial fraction, or 
mix at all: Displacement of fluid parcels of different $Z$ by the
velocity field conserves their volumes and thus volume fractions (or 
mass fractions for compressible flows), replacing one by another in space,
having no effect on the $Z$ distribution. 
This can be shown rigorously using a metallicity pdf equation (PS). 
A velocity field can only enhance
mixing by spatially ramifying the $Z$ field for diffusion to operate 
on small
scales. 
Models that mix by sweeping of
gas by SN or SB shells, cloud motions, differential 
rotation,
or ``turbulent diffusion'' are unphysical without recognition that they are
implicit models for microscopic diffusion.
\subsection{Several Catalyzing Velocity Fields}

Expanding SNRs and superbubbles (SBs) are the main agents of mixing in many
sequential enrichment inhomogeneous chemical evolution models 
(Reeves 1972; see Tsujimoto et al. 1999;
Oey 2000; Argast et al. 2000; Saleh et al. 2006)
\nocite{ree72, tsu99, oey00, arg00, sal06}.  Shells can mix, but only if
instabilities allow diffusion to mix swept-up gas with new products of
nucleosynthesis. Assuming this occurs, each SNR sweeps up and mixes a
mass $M_{sw} \sim 2 \times 10^4$ $M_\odot$ for $Z=0$ gas \citep{tho98, 
han06}.
The time to sweep up primordial ISM is $\tau_{sw} =
M_{gas}/ (\nu_{SN} M_{sw})$, where the SN rate 
$\nu_{SN}= \epsilon B/\langle M_* \rangle$. 
For an IMF with mass range (0.1, 100) $M_{\odot}$ and indices (-0.4, -1.7) 
below and above 1 $M_{\odot}$ the number fraction $\epsilon$ of stars that 
become SNe is 0.004 and the average stellar mass
$\langle M_* \rangle=0.6$, so $\nu_{SN} = 0.7/yr$ and 
$\tau_{sw}=3 \times 10^6$ yr. 
Equivalently, the accumulated volume filling factor
$NQ = t/\tau_{sw}$ \citep{oey00}, so
$P= exp (-t / \tau_{sw})=exp(-NQ)$ and it is impossible to
preserve primordial gas in LBGs for longer than $\sim 1\% $ of the
observed SF ages $\sim 3 \times 10^8$ yr, unless the mixing 
efficiency is
artificially tuned to 1\%, in which case the model predicts too large a
present-day scatter in metallicity compared to observations. 
Superbubbles
have a smaller frequency, but sweep
up more gas, producing an almost identical result. Spatial clustering 
and
infalling $Z = 0$ gas also do not affect the conclusion. These points
are discussed in details in a separate publication (PS).
Using the same argument as for $\tau_{src}$ in sec 2.2, an IMF cutoff
at 1 $M_{\odot}$ does not affect the SN rate, giving the same
$\tau_{sw}$. Unexpectedly, $\tau_{sw}$ for a VMS IMF is also nearly
unchanged, because of the cancellation of the decrease of $\nu_{SN}$ 
(by a factor
of 100) and the increase of $M_{sw}$ by a factor of $\sim$ 50 due to 
the large
explosion energy of pair instability SNe (up to $10^{53}$ erg, see 
Woosley et al. 2002).
Therefore our conclusion that SN or SB sweep-up mixing is so fast 
that LBGs should have zero primordial gas is virtually independent of 
the IMF. The mixing would be 100 times faster using the mixed mass per event
adopted by Tumlinson (2006).

Unmixed pockets of metals in SBs could blast out of a galactic disk,
later showering the disk with ``droplets'' of pure metals,
diffusively mixing once they land in the disk \citep{ten96}.
Fine-tuning of the number of
droplets, or equivalently the mixed mass, per SN, is required
to give the desired timescale, but is unspecified by the model. 

Differential rotation could stretch the products of nucleosynthesis 
deposited
in a $\sim 100$ pc SN blob, into long thin annuli until the scale of
diffusivity is reached. The shear rate in LBGs, or whether they
differentially rotate, is unknown. We used the rate $\sim 10$ km/sec/kpc
in our Galaxy as an illustration and found that the timescale to reach
the diffusive scale derived in sec. 3 below, is about $8$ Gyr, which
is too slow.


Another stretching process, turbulence, produces exponential, 
rather than
linear, stretching, with strain rates ten
times larger than Galactic differential rotation.

\section{Turbulence-Enhanced Diffusive Mixing}
Turbulence deforms large-scale fluid elements and the tracers
they contain into filaments and sheets, bringing tracers closer
together until scales are reached on
which diffusivity can homogenize faster than the strain timescale of the
fluid. The process is described by the general equation for the 
evolution
of the metallicity field in an arbitrary velocity field ${\bf u}({\bf  
x}, t)$,
\begin{equation}
\frac{\partial Z({\bf x}, t)} {\partial t} + {\bf u}({\bf x}, t)\cdot 
{\bf \nabla} Z({\bf x},t) = \frac{1}{\rho} \nabla \cdot (\rho \kappa \nabla 
Z({\bf x},t))+S
\end{equation}
where $\kappa$ is the diffusivity and $S$ denotes the sources.

In our model a straining
event on the scale of the sources $L_s$ takes $Z$ to a critical scale,
$L_{diff}$, small enough for diffusivity to operate, at constant
mean strain rate in a single step, by exponential stretching of line 
elements
\citep{bat52,vot02}. This ``short circuit'' of the scalar cascade
\citep{vil01} is supported by experiments \citep[]{vil04,vot02} and
simulations \citep{gir90, got03}, 
and is similar to the
scalar turbulence field theory of Shraiman and Siggia (2000).\nocite 
{shr00}
We assume that in supersonic turbulence compressions are analogous
to stretching in the sense of bringing tracers to the
critical scale $L_{diff}$.

The scale $L_{diff}$ below which the diffusivity term exceeds the 
advection term
in eq 1 is obtained by equating the two terms and replacing
spatial derivatives by $L_{diff}^{-1}$ and ${\bf u}$ by the velocity at
scale of $L_{diff}$. Assuming $u_l \sim l$ scaling appropriate for
exponential stretching, the result is
$L_{diff} = [\kappa/ (U/L_s)] ^{1/2}$, where $U$ is
the rms turbulent velocity on the scale of the sources.
A residence-time average diffusivity, assuming the WNM contains more
than 20\% of the ISM mass, is $\kappa \sim 10^{20}$ $cm^2 s^{-1}$. Then 
the
average diffusivity scale in the ISM is
$L_{diff} \sim 0.06$ $(L_{100}/U_{10}) ^{1/2}$ pc where
$L_{100}$ is $L_s/100$ pc, and $U_{10}$ is U in units of $10$ km/s 
\citep[]{kul87}. We assume the same scale for $z \sim 3$ LBGs,
noting that $L_{diff}$ only enters the mixing time (below) 
logarithmically.

The mixing time follows from the assumed exponential stretching, in 
which
scales change as $dl/dt = -\gamma l$, where $\gamma = U/L_s$ is the
large scale strain rate. This gives a time to
bring tracers from $L_s$ to $L_{diff}$ as $\tau_{mix}$ =
$(L_s/ U)$ $ln(L_s/L_{diff})=(L_s/2U)$ $ln(UL_s/\kappa)$.
The quantity $UL_s/\kappa$ is the diffusivity analogue of the 
Reynolds number,
called the Peclet number $Pe$. Numerically
$Pe \sim 3 \times 10^7$ $(U_{10} L_{100}/\kappa_{20})$, giving
$\tau_{mix} = (L_s/2U)$ $ln (Pe) \sim 75$ Myr. The result could be
smaller if the velocity dispersion increases with the SFR, as for
turbulence driven by SNe \citep[e.g.,][] {dib05}.

The primordial fraction is the cumulative pdf of the gas metallicity
$\int_0^{Z} f(Z',t) dZ'$, where $f(Z,t)$ is the differential pdf of
metallicity, as the integration limit $Z$ approaches zero. We
can obtain the equation for $P(t)$ heuristically, without details of
the integral closure \citep{jan79} we used to derive the full pdf 
equation corresponding to the advection-diffusion equation eq 1 (PS), a
turbulent mixing closure that gives the same timescale for exponential 
variance decay as found in simulations by de Avillez and Mac Low 
(2002, eq. 17), and the same dependence of mixing time on initial size and SN 
rate (their Fig.7), if velocity dispersion $U$ scales as the square root of 
the SN rate (Dib and Burkert 2005).

The primordial fraction decreases whenever fluid elements 
with primordial mass fraction $P$ and gas that has been
polluted by sources or previous mixing events, with mass fraction $1-P$,
are stretched
sufficiently to result in diffusive mixing. This interaction occurs 
with an
average frequency $\tau_{mix}^{-1}$. $P$ is also
gradually depleted by cycling through massive stars that inject 
metals
when they explode, on a timescale $\tau_{src}$(sec 2.2), 
which we
assume is constant for the times of interest. The equation for the
primordial fraction is then
\begin{equation}
dP/dt = -P(1-P)/\tau_{mix} - P/ \tau_{src}
\end{equation}
whose solution is, using the fact that $\tau_{mix} \ll \tau_{src}$,
\begin{equation}
P = (1 + (\tau_{mix}/\tau_{src}) exp( t /\tau_{mix}))^{-1}
\end{equation}

\begin{figure}
\centerline{
\epsfxsize=8cm \epsfbox{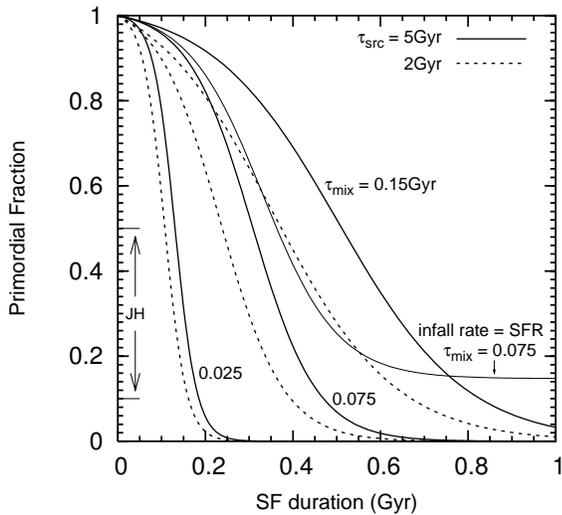}
}
\caption[]{Primordial fraction as a function of SF duration for 
combinations of the mixing timescale $\tau_{mix}$ and the source
timescale $\tau_{src}$ (see sec 2.2).
The empirical value for $\tau_{src}$ is about
5 Gyr within a factor of a few for an IMF with $M_l=0.1$. 
$\tau_{src}$ is independent of $M_l$ 
if it is smaller than the mass limit for SNe. 
For a VMS IMF, $\tau_{src}$ is much larger but does not
affect the result significantly, as discussed in the text.
The range suggested for $z \sim 3$ LBGs (JH) is
indicated by the arrow. Also shown is an infall model with infall rate
equal to the SFR.\label{f1}}
\end{figure}

$P$ decreases exponentially on timescale $\tau_{mix}$, but
only after a delay time $\sim \tau_{mix}$ $ln (\tau_{src}/\tau_{mix})$,
which is $\sim 3 \times 10^8$ yr for $\tau_{mix}=75$ Myr and
$\tau_{src}=5$ Gyr assuming an IMF with $M_l \lsim 8$ $M_{\odot}$.
The delay time is the time for sources to provide enough
non-primordial gas to make $P$ depart from unity, but the dependence
on $\tau_{src}$ is only logarithmic, and so is nearly independent
of IMF, even for the
extreme case of a VMS IMF (see sec 2.2).

The behavior of $P$ as a function of time for different
$\tau_{mix}$ and $\tau_{src}$ is illustrated in Fig. 1.
The smallest mixing timescale shown, 0.025 Gyr, corresponds to a 
larger velocity dispersion
$\sim 30$ km/s, not unreasonable for galaxies with large SFRs. Our major
result is that $P(t)$ declines on a timescale similar to SF ages 
inferred from empirical modeling, without adjustment of parameters.

The effect of $Z=0$ infall can be understood by adding to eq 2 a term
$(1-P)/\tau_{in}$ where $\tau_{in} =M_{gas} /$infall rate is the infall
timescale. An example is shown in Fig. 1. Infall allows
intermediate values of $P(t)$ for a longer time, but only for infall timescales
close to $\tau_{mix}$, implying a
huge infall rate. Therefore it is unlikely that infall modifies the
primordial fraction predicted by turbulent mixing. Galactic winds with
large rates \citep{erb06} have no effect
if the winds sample the full pdf of metallicity. If the galaxies have 
undergone previous episodes of SF with an accumulated
duration as large as  $\sim 1$ Gyr, even the turbulent model cannot explain
the intermediate primordial fractions claimed by JH.

\section{Discussion}
Most mixing processes predict a
primordial fraction that is either unity or zero at $z \sim 3$ because
they mix on a timescale that is much larger or smaller
than the empirical SF ages. 
$P(t)$ should be zero in almost all galaxies if stellar explosions 
mix as efficiently as assumed in sequential enrichment models (or much more
efficiently, Tumlinson 2006). 
Our turbulence-enhanced diffusivity model 
naturally preserves primordial gas from rapid mixing for a few times the 
mixing time, which itself depends only weakly
on parameters, in particular the assumed 
IMF or the averaged diffusivity, and gives an intermediate primordial 
fraction in galaxies with SF ages $\sim 1-3 \times 10^8$ yr.
That this timescale happens to match the star
formation ages of these galaxies is no coincidence if star
formation is driven by turbulence \citep{mac04} powered by stellar
explosions.
Future systematic investigations of the spectrophotometric signatures
of primordial gas in galaxies could distinguish these possibilities.

\acknowledgements
We thank J. Craig Wheeler and the referee for constructive comments.  
This work was supported by NASA ATP grant NAG5-13280.


\end{document}